\documentclass[12pt]{article}
 
 \newcommand{\n}{\noindent} 
\newcommand{\rf}[1]{(\ref{#1})}
\newcommand{\ba}{\begin{array}} \newcommand{\ea}{\end{array}}
\newcommand{\be}{\begin{equation}} 
\newcommand{\btb}{\begin{tabular}}\newcommand{\etb}{\end{tabular}}
\newcommand{\ee}[1]{\label{#1}\end{equation}}
\newcommand{\bi}{\bibitem} 
 
\newtheorem{thm}{Theorem}[section] \newtheorem{pro}[thm]{Proposition}
\newtheorem{df}[thm]{Definition} 
\newcommand{\dss}{\displaystyle}
\newcommand{\bfl}{\begin{flushleft}}\newcommand{\efl}{\end{flushleft}}

\newcommand{\al}{\alpha} \newcommand{\bt}{\beta}
  
\newcommand{\ep}{\epsilon}
\newcommand{\te}{\theta} \newcommand{\De}{\Delta}
\newcommand{\la}{\lambda}  \newcommand{\si}{\sigma}
\newcommand{\La}{\Lambda}
\newcommand{\Si}{\Sigma}
\newcommand{\C}{\mathbb C}  \newcommand{\R}{\mathbb R}

\newcommand{\LCU}{{\cal L}_\uparrow}
\newcommand{\PCU}{{\cal P}_\uparrow}
\newcommand{\PC}{{\cal P}}

\newcommand{\TC}{{\cal T}}
\newcommand{\GC}{{\cal G}}\newcommand{\hGC}{\hat{\cal G}}
\newcommand{\HC}{{\cal H}}

\newcommand{\0}{{\bf 0}}
\newcommand{\by}{{\bf b}}

\newcommand{\bty}{\mbox{\boldmath $\beta$}}

\newcommand{\ey}{{\bf e}} 
 
 \newcommand{\my}{{\bf m}}
 
\newcommand{\ky}{{\bf k}}\newcommand{\ny}{{\bf n}} 
\newcommand{\ry}{{\bf r}}

\newcommand{\Vy}{{\bf V}}
 
\newcommand{\uy}{{\bf u}}

 \newcommand{\we}{\wedge} 
\newcommand{\ra}{\rightarrow}

\newcommand{\Bi}{\mbox{Bi}}

\newcommand{\vers}{\mbox{vers}}

\usepackage{latexsym}
\usepackage{amssymb}
\usepackage[T1]{fontenc}
\usepackage{graphicx}

\begin{document} 

\title{Spatial Directions, Anisotropy and Special Relativity}
\normalsize
\author{Marco Mamone Capria\\ \small Dipartimento di Matematica -- via Vanvitelli, 1 -- 06123 Perugia -- Italy \\
\small {\sl E-mail}: mamone@dmi.unipg.it}
\maketitle

\begin{quote}\small
{\bf Abstract} The concept of an objective spatial direction in special relativity is investigated and theories assuming light-speed isotropy while accepting the existence of a privileged spatial direction are classified, including so-called very special relativity. A natural generalization of the proper time principle is introduced  which makes it possible to devise non-optical experimental tests of spatial isotropy. Several common misunderstandings in the relativistic literature concerning the role of spatial isotropy are clarified.


\end{quote}\normalsize

\section{Introduction}

The link between the law of constancy of the speed of light and the Poincar\'e group $\PC = \PC (4)$, via the principle of relativity, is not so direct as is sometimes described in the textbooks.

In fact there is nothing in the principle of relativity plus the constancy law to forbid selecting a subgroup of the proper orthocronous conformal Poincar\'e group $C\PCU^+$ which is not contained in $\PCU^+$. The further ingredient that must be added in order to forbid this result is an isotropy assumption. This has been realized, to some extent, since the beginnings of special relativity (\cite{poi}, \cite{ei05}).

However misunderstandings on this topic still exist. For instance, in a book on the foundations of special relativity the choice of a constant Reichenbach function $\ep =1/2$ (cf. \cite{mm}) is defended on the ground that ``the main insight of the Special Theory" is that ``there are no preferred frames of reference, {\sl and hence no preferred directions}'', which unduly conflates the isotropy {\sl of the one-way velocity of light} with the isotropy {\sl of space} (\cite{lh}, p. 61; italics added). Or, to give another example, a classical treatise qualifies the introduction of spatial isotropy (without even naming it) as ``a simple relativity argument'', which is surely much too hasty (\cite{mo72}, pp. 36, 37). To emphasize that spatial isotropy can be empirically tested, whatever one thinks of the one-way velocity of light, is one of the aims of this paper, which is meant to offer a logically transparent path to the anisotropic special relativity, the variant of special relativity introduced in Bogoslovsky's pioneering contributions (\cite{b}, \cite{b92}, \cite{bg}, \cite{gb99}). 

The main idea is to elaborate the anisotropy assumption  by exploiting the concept of an objective spatial direction in conformal Minkowski space-time, and to use it in a very natural way to select a subgroup of the conformal Poincar\'e group different from the Poincar\'e group itself. In other words, starting with a space-time having as a structure group the conformal Poincar\'e group one can perform a reduction of this group in a different way from the usual one. 

We shall see that a suitable generalization of the proper time principle makes it possible to derive uniquely the Finsler metric which takes the place of the Lorentzian metric and to design a test for spatial isotropy based on differential aging (the classic `twin paradox'). The link with the proposal of a version of special relativity with a structure group smaller than the Poincar\'e group (so-called ``Very Special Relativity'') is also elucidated. In the final section it will be stressed that spatial isotropy is not equivalent to reciprocity and should be kept distinct from the isotropy of the one-way velocity of light (or light-speed isotropy); on the other hand, it is equivalent (at the level of special transformations) to the famous `symmetry' of the elementary relativistic effects that many authors wrongly consider as indissolubly linked with the very idea of `relativity'.

I tried to make this article reasonably self-contained, because I think that the present topic deserves much more attention, on both theoretical and pedagogical grounds, than it is usually allotted in the textbooks. As to the pedagogical side, it is my opinion that a good understanding of the derivation of the special Lorentz transformations requires considering several of the issues dealt with in this paper (at least those needed to understand sections 2.1.1 and 5 ). I am not aware of a single textbook on relativity which tries to explain that spatial isotropy is not a trivial or necessary ingredient of the principle of relativity, by at least {\sl exhibiting} the special anisotropic transformation (that is, the system of equations \rf{tsla}). I hope this paper may help that to change.

\section{Basic definitions}

In this section we recall briefly a number of essential notions, mainly in order to fix our notation and usage of terms.

A Lorentzian form on a real vector space $V$ (of at least 2 dimensions) is a symmetric bilinear form with signature $(+,\dots,+,-)$. The corresponding Lorentzian norm is the map

\[ |\cdot|: V\ra \R^+, \; v\mapsto \sqrt{|g(v,v)|}. \]

A Lorentzian vector space is a pair $(V,g)$ where $g$ is a Lorentzian form on $V$; a conformal Lorentzian vector space is a pair $(V, \R^+ g)$, which means that $g$ is determined up to a positive factor.

In a Lorentzian vector space $(V, g)$ the spacelike vectors are the zero vector and those $v\in V$ such that $g(v,v)>0$; the timelike vectors fill the open set $\TC$ defined by the inequality $g(v,v)<0$; the lightlike (or null) vectors fill the light-cone, that is the set $C$ of all nonzero vectors $v\in V$ with $g(v,v)=0$. Both $\TC$ and $C$ are disconnected sets, with two (path) connected components. If we select one of the components of $\TC$, we are fixing a time orientation for $(V,g)$; the vectors belonging to the selected component (resp. to the other component), denoted by $\TC^+$ (resp. $\TC^-$), are the future-pointing (resp. past-pointing) timelike vectors. The same definitions apply to the conformal case.

A Minkowski vector space is an oriented and time-oriented Lorentzian 4-dimensional vector space (or, generally, of at least 2 dimensions). A Minkowski space-time is a real affine space $M$ such that its associated vector space of translations $V= V(M)$ is a Minkowski vector space. The conformal variants of both notions are defined in the obvious way. 

In general, if $\GC$ is a subgroup of the group of bijections of $\R^n$ to itself, and $X$ is any set with the cardinality of continuum, then a $\GC$-structure on $X$ is an orbit of the set $\Bi (X,\R^n)$ of all bijections from $X$ to $\R^n$ with respect to the natural (left) action of $\GC$. A pair of the form $(X,\Phi)$ where $\Phi$ is a $\GC$-structure on $X$ is a $\GC$-space and the elements of $\Phi$ are its admissible coordinate systems. 

With this definition in mind, we can consider equivalently a Minkowski space-time to be a $\PCU^+$-space, where $\PCU^+$ is the proper orthochronous Poincar\'e group, that is $\PCU^+ : = \LCU^+ \rtimes \R^4$, and $\LCU^+$ is the proper orthochronous Lorentz group, namely the group of all $4\times 4$ real matrices $\La = (\la_{ij})$ such that:

\[ \La^T G \La = G, \; \la_{44} >0, \det \La>0, \]

\n
where $G$ is the diagonal matrix with entries $(1,1,1,-c^2)$. 

A conformal Minkowski space-time is a $C\PCU^+$-space, where $C\PCU^+ : = \R^+\LCU^+ \rtimes \R^4$, is the conformal proper orthochronous Poincar\'e group, a 11-dimensional Lie group.\footnote{Some authors call it the `Weyl group' (see e.g. \cite{ggp}).} 

A time orientation determines a distinction also between future-pointing (resp. past-pointing) lightlike (or null) vectors according to the topological condition (``Bd''  denotes here the topological boundary):

\[ C^{\pm} = \mbox{Bd} (\TC^{\pm})\setminus\{0\}.\]

An inertial observer in Minkowski space-time is a vector $u\in\TC^+$ with Lorentzian norm \( |u| = c\). Given an inertial observer $u$, its  (standard) simultaneity (vector) space is the orthogonal subspace $S_u = [u]^{\perp}$. The pair formed by $[u]^\perp$ together with the corresponding restriction of $g$ is an Euclidean 3-space. Given an inertial observer $u$ and an origin (that is, a selected event $o\in M$) one can construct infinitely many Minkowski coordinate systems $\phi: M\ra \R^4$, one for every choice of an orthonormal basis $(e_1, e_2, e_3)$ of $S_u$ such that the basis $(e_1, e_2, e_3, u)$ is positively oriented ($\phi (p)$ is simply the 4-tuple of components of the vector $p-o$ in the latter basis). A basis of the form $(e_1, e_2, e_3, u)$ just described is called a Minkowski basis.

Every inertial observer $u$ defines two orthogonal projections according to the decomposition of $V$ as the direct sum 
$[u] \oplus S_u$, which allows us to write uniquely for every vector $v$ the equality  \( v= v^{||} + v^\perp\). We shall denote also as

\[ P_u : V \ra S_u, \; v \mapsto v^{\perp} = v+\frac{g(v,u)}{c^2} u , \]

\n
the map taking each vector to its spatial component with respect to $u$.

A famous fact in special relativity is that different inertial observers define different simultaneity spaces, contrary to what we are used to in classical physics.\footnote{Some authors consider the standard correspondence between inertial observers and simultaneity spaces as merely conventional; see \S 5.3.}   This is important for our purposes, because it implies that there is nothing like a `spatial vector' which can be talked about independently of the choice of an inertial observer. Thus to speak of {\sl spatial} isotropy, or of its absence, without absolute simultaneity is necessarily more tricky than in classical physics. The good news is, however, that there exists a valid observer-independent notion of a {\sl spatial direction}, which can be identified across different simultaneity spaces.

\subsection{Spatial directions}

If $w$ is a nonzero vector in $S_u$, there is exactly one future-pointing lightlike vector $\ell$ such that $w$ is the spatial component of $\ell$ with respect to $u$; explicitly, if we denote this vector by $\ell^{[u]}_w$, we have

\[ \ell^{[u]}_w = w +\frac{|w|}{c} u. \]

\n
If $u,u'$ are inertial observers, then the spatial direction of any nonzero vector $w$ in $S_u$ can be identified with the direction of $P_{u'}\ell^{[u]}_w$ in $S_{u'}$. By construction we have that:

\[ \ell^{[u']}_{P_{u'}(\ell^{[u]}_w)} = \ell^{[u]}_w. \]

\n
The map from $S_u$ to $S_{u'}$ sending $w\mapsto P_{u'}\ell^{[u]}_w$ is therefore a bijection (if we extend it by sending the zero vector to itself) and is positively homogeneous, but it is not linear, so it is not an isomorphism of vector spaces, while the restriction of $P_{u'}$ to $S_u$ of course is. However, it is enough for the spatial directions to be parametrized by half-lines in $C^+$, in a way that does not depend on the inertial observer; notice also that the whole procedure can be used in a conformal Minkowski vector space as well. Formally we give the following definitions.

\begin{df} A {\bf ray}  in Minkowski space-time (or in conformal Minkowski space-time) is a positive half-line generated by a future pointing light-like vector.\end{df} 

Note that in the (1+1)-dimensional Minkowski space-time, so much liked by both textbook authors and workers in the foundations of relativity, there are only {\sl two} rays, which oversimplifies the whole issue of spatial isotropy. In fact the space of rays in a 4-dimensional Minkowski space-time is easily seen to be topologically equivalent to a 2-sphere. So although in special relativity (as normally interpreted) there is not a single 3-dimensional space for all admissible coordinate systems, yet all admissible coordinate systems share the same celestial sphere.\footnote{A simple but instructive discussion about the extent this `sameness' holds can be found in \cite{pen07}, pp. 428-30.} 

\begin{df} Let $u,u'$ inertial observers; two nonzero vectors $w\in S_u, w' \in S_{u'}$ {\bf have the same spatial direction} if $\ell^{[u]}_w$ and $\ell^{[u']}_{w'}$ generate the same ray.\end{df}

We denote $L^{[u]}_w : = \R^+ \ell^{[u]}_w$, so $w\perp u$ and $w'\perp u'$ determine the same spatial direction with respect to the inertial observers $u$ and $u'$, respectively, if and only if $L^{[u]}_w = L^{[u']}_{w'}$.

Let $b = (e_1, e_2, e_3, u)$ and $b'= (e'_1, e'_2, e'_3, u')$ Minkowski bases, call $\La\in \LCU^+$ the matrix connecting them according to $b = b'\La$), and suppose that they give the same homogeneous coordinates to the ray $L = \R^+ \ell$, that is, $\ell = bl = \la b' l$, with $l\in \R^4$; we have immediately that 

\be\La l = \la l.\ee{eilo} 

\n
Clearly we can take, with no loss of generality, $l$ to be of the form

\[ l = (\ny, 1/c), \, |\ny| =1 . \]

\n
Using from now on the representation of the Lorentz matrix described in the Appendix and taking the 4-th component on both sides of \rf{eilo}, we obtain

\be \la = \al (1-\bty\cdot\ny), \ee{lafac}

\n
and for $S\in SO(3)$ such that $\La = \Si_S \La (\Vy)$ we have:

\be S(\ny + (\al -1)(\uy\cdot \ny)\uy - \frac{\al}{c}\Vy) = \al (1-\bty\cdot\ny)\ny .\ee{eilos} 

\n
We are now going to state a result, which characterizes up to a 1-parameter subgroup of rotations the special Lorentz transformations in terms of the concept of spatial direction. First we remember that if $\Vy = \Vy_\La$ is the velocity of the Lorentz group matrix $\La$ giving the linear part of the transition function from a Minkowskian coordinate system $\phi$ to another one (or the matrix between the corresponding Minkowski bases, in the right order), then the {\sl reciprocal velocity} is defined as $\Vy_{\La^{-1}}$. It is easy to check that if we consider any inertial tranformation  $x'= Bx + b$ where

\be B = \left(\ba{cc} A & -A\Vy \\ \ky^T & \al \ea\right), \ee{block}

\n
then the condition that the velocity of $\phi$ with respect to $\phi'$ (which is $-\al^{-1} A\Vy$) is the opposite of $\Vy$ is equivalent to \( A\Vy =\al\Vy  \), and for a Lorentz transformation $\La$ this means that $(\uy, 1/c)$, $\uy$ being the versor of $\Vy$, is a (null) eigenvector (with eigenvalue $\al (1-\bt)$) of $\La$. 

\begin{pro} Let the Minkowski bases $b, b'$ be related by $\La = \Si_S \La (\Vy)$, according to the equation $b = b'\La$, and let $l = (\uy, 1/c)$, where $\uy$ is the versor of $\Vy$. 

1) The following conditions are equivalent: (i) $\Vy$ is the opposite of the reciprocal velocity; (ii) $S\Vy = \Vy$; (iii) $l$ is an eigenvector of $\La$; (iv) $b$ and $b'$ give the same homogeneous coordinates to the ray defined by $l$. 

2) If, in addition to (one of) the conditions (i)-(iv), $b$ and $b'$  give the same homogeneous coordinates also to another, non parallel ray, then $b=b'$.\end{pro} 

\n
{\bf Proof} 1) The proof has been partly anticipated and for the rest it is a straightforward verification. Notice that by substituting $\uy$ for $\ny$ in \rf{eilos} we obtain $S\uy =\uy$; the reverse is also easy to see by direct computation. 

2) Let $l_1 = (\ny, 1/c)$ be another null vector whose ray has the same homogeneous coordinates according to both $b$ and $b'$, then we can apply \rf{eilos} to it, and obtain:

\[ S\ny + ((\al -1) (\uy\cdot\ny) - \al\bt)\uy = \al (1-\bt \uy\cdot\ny) \ny. \]

\n
Taking the scalar product of both sides with $\uy$, and using the fact that $S\ny\cdot\uy = S\ny\cdot S\uy = \ny\cdot\uy$, we obtain, after a few easy passages: \(\al\bt (\uy\cdot\ny)^2 = \al\bt\) and therefore $\bt= 0$ (since $\ny \neq \pm\uy$ by assumption). By substituting $\Vy =\0$ in the previous formula we obtain $S\ny = \ny$; now a rotation in $\R^3$ fixing two independent vectors can only be the identity, and the claim follows.\hfill $\Box$  

\subsubsection{On diagrams illustrating the Lorentz transformation}

The preceding proposition shows that the standard diagrams illustrating {\sl in 3-dimensional terms} the special Lorentz transformation, with `parallel' homologous spatial axes (see Figure 1), should not be taken seriously and can easily mislead the non-initiated, as they suggest that at least in some special cases reference frames having a nonzero relative velocity may nonetheless have the same three spatial directions for the homologous axes. Probably it is this misperception of the Lorentz transformation  that makes it so difficult to many students (and others!) to see that the Thomas precession is not a logical absurdity.\footnote{``It took Pauli a few weeks before he grasped Thomas's point'' (\cite{pai82}, p. 144). Notice that this happened in 1926, that is 5 years after the young Pauli had published his article ``Relativit\"atstheorie'', that was to become a famous reference work.}

\begin{figure}[htp]
\centering
\includegraphics[totalheight=0.3\textheight ]{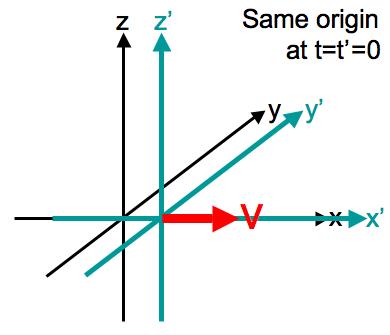}
\caption{A typical textbook diagram, from Wikipedia's ``Lorentz transformation''}
\end{figure}

Incidentally, the identification of the space of rays with the Riemann sphere provides a quick confirmation of the last statement (which is a variant of point 2) of the proposition 2.3). By using the universal 2:1 covering 

\[ SL (2,\C) \ra \LCU^+ \]

\n
we have that every Lorentz transformation defines a unique automorphism of the Riemann sphere. Now, every automorphism of the Riemann sphere can be represented as a M\"obius transformation, that is a complex function of the form 

\[ f(z) = \frac{az + b}{dz + f}, \; af-bd =1 \]

\n
and it is easy to see that only the identity fixes three points. This shows that any Lorentz transformation preserving three spatial directions is necessarily the identity.

\section{Privileged spatial direction}

In this section we will discuss how to implement within the framework of the principe of relativity, the notion that physical laws are the same in a class of coordinate systems giving the same homogeneous coordinates to a privileged spatial direction. We shall treat the classical case first. 

\subsection{Classical physics}

In {\sl classical physics}, by which I mean here the relativistic physics having the Galileo group $\GC_G$ as its invariance group, the condition that the direction defined by $\ny$, a unit vector in $\R^3$,  is privileged can be expressed as follows: all pairs of admissible coordinate systems $\phi, \phi'$ which a priori have a transition function $\phi'\circ\phi^{-1}$ of the form

\[ x'= B x+ b, \;\mbox{with}\;  B=\left(\ba{cc} S & -S\Vy \\ 0 & 1 \ea\right), \; S\in SO(3), \; b\in\R^4. \]

\n
are subjected to the further condition that $(\ny, 0)$ be an eigenvector of $B$. This is equivalent to $S\ny = \ny$, which selects a subgroup isomorphic to $SO (2) \cong U(1)$. The {\sl anisotropic Galileo group} is therefore:

\[ \GC_{AG} = (SO(2) \rtimes R^3) \rtimes \R^4. \]

\n
and has dimension 8, that is two less than the original Galileo group. This gives the right indication also for the special relativistic case.

\subsection{From conformal Poincar\'e group to the anisotropic Poincar\'e groups}

In special relativity, the invariance of the speed of light and the inertia principle leads to a general affine transition function between admissible coordinate systems $\phi, \phi'$ of the form:

\[ x'= \mu\La x+ b, \;\mbox{with}\; \mu\in\R^+, \; \La\in\LCU^+, \; b\in\R^4. \]

\n
Here $\La$ varies a priori in a subset of the proper orthocronous Lorentz group $\LCU^+$. A priori the conformal factor $\mu$ is to be regarded as a function $\mu(\phi,\phi')$. If, furthermore, we take 1) $\mu$ to be a function of $\La$ and 2) its domain to be the whole of $\LCU^+$, then it follows that $\mu$ is identically equal to 1. However, if we remove 2), this conclusion is no more warranted. As a matter of fact, the isotropy of the one-way velocity of light (taken together with a maximality condition on the group) leads more directly to the {\sl conformal} Minkowski space-time than to Minkowski space-time itself. On the other hand, if $C\PCU^+$ is too large as a structure group, there is more than one way to achieve a reduction.

In fact suppose that there is a {\sl privileged spatial direction}, which we can describe as a ray $L$. If the components of a generator of $L$ in a given coordinate system $\phi$ are given by the 4-tuple $l$, then all the other admissible systems will be those $\phi'$ such that the linear part of the transition function $\phi'\circ\phi^{-1}$ has $l$ as an eigenvector with some eigenvalue $\la$, which in principle depends on $\La$ and $\mu$:

\be (\mu\La)l = \la l. \ee{anis}

\n
This equation just means that $\La$ maps the ray $\R^+ l$ in the numerical Minkowski space-time $\R^4$ to itself. 

It is instructive to discuss equation \rf{anis} without assuming from the start that $l (\neq 0)$ is lightlike.\footnote{This is similar to the approach followed by Wigner in his famous 1939 paper, where he introduced his ``little groups'' (\cite{wig}). The paper has been reprinted in a useful collection, with commentaries (\cite{nk}).} First restrict the eigenvalue function to be $\la \equiv 1$:

\be (\mu_0 \La)l = l. \ee{anisp}

\n
It is clear that \rf{anisp} is a group-defining condition; we will denote by $\hGC_0$ the subgroup of the conformal  group $C\LCU^+$ corresponding to \rf{anisp}, that is:

\be \hGC_0 = \{ \mu_0 \La\in C\LCU^+ \; : \; \mu_0 \La l = l\}. \ee{g_anisp} 

\n
If for $\mu_1 , \mu_2 \in\R^+$ both $\mu_1\La$ and $\mu_2\La$ belong to $\hGC_0$, then$(\mu_1 - \mu_2)\La l = 0$, which implies that  $\mu_1 = \mu_2$ since $\La$ is nonsingular. In other terms, $\mu_0$ {\sl is a function of} $\La$; more precisely, if we put

\[ \GC_0 : =\{\La\in\LCU^+ \; : \; \exists \mu \in\R^+ \; \mbox{s.t.}\; \mu \La\in\hGC_0\}, \]

\n
we have that also $\GC_0$ is a group and $\mu_0 : \GC_0 \ra \R^+$ {\sl can be regarded as a group homomorphism}, so that we can also describe $\hGC_0$ as:

\[\hGC_0 = \{\mu_0 (\La)\La : \; \La\in \GC_0 \} = \GC_0\times_{\mu_0} \R^+, \]

\n
where $\mu_0$ is the map defined by \rf{anisp}. Notice that from \rf{anisp} it follows that

\[ g_c (l,l) = g_c ((\mu_0 \La)l, (\mu_0 \La)l) = \mu_0^2 g_c (l,l) \]

\n
so the only possibility for $\hGC_0$ not to be a subgroup of $\PCU^+$ (that is, for $\mu_0$ not to be identically 1) is for $l$ to be a lightlike vector: $g_c (l, l) = 0$. 

\begin{df} The group $\GC_0 \equiv \GC_0 (L)$ (resp. $\hGC_0 \equiv \hGC_0 (L)$  is the {\bf anisotropic Lorentz group} (resp. {\bf anisotropic conformal Lorentz group}) for the ray $L = \R^+ l$. The semi-direct product of $\HC$ with the translation group of $\R^4$, namely \(\HC \rtimes T(\R^4)\), is the {\bf anisotropic Poincar\'e group} $\GC$ if $\HC = \GC_0$ and the {\bf anisotropic conformal Poincar\'e group} $\hGC$ if $\HC = \hGC_0$.\end{df} 

\n
{\bf Remark} It is easy to check that condition \rf{anisp} for timelike $l$ singles out a subgroup of $\LCU^+$ which is isomorphic to $SO(3)$ (or, equivalently, to the group $SO_4 (3)$ of the spatial rotations in Minkowski $\R^4$, see Appendix) while if $l$ is a nonzero spacelike vector, then  the resulting subgroup is isomorphic to $\LCU^+ (3)$. Together with $\GC_0$ these are Wigner's ``little groups" (\cite{wig}).

\subsection{Basic properties of the anisotropic Lorentz groups}

As in section 2, we take $L$ to be generated by $l = (\ny, 1/c)$ with $|\ny|=1$ (of course this implies no loss of generality).

First remark that $\GC_0$ is surely nontrivial, because it contains all rotations fixing $\ny$, that is, all matrices $\Si_S$ with  $S\ny = \ny$; clearly for such matrices it is $\mu_0 (\Si_S) =1$. The group $\GC_0$ also contains the special Lorentz transformation $\La(\bty)$ with $\bty$ proportional to $\ny$; in this case $\mu = \al (1\pm\bt)$, where $\bt := |\bty|$. In general if we represent $\La$ as a block matrix (see Appendix for details):

\be \La = \left(\ba{cc} A & -c A\bty \\ -\frac{\al}{c}\bty^T & \al \ea\right), \ee{blocks}

\n
equation \rf{anisp} translates into

\be A(\ny - \bty) = \frac{\ny}{\mu}, \;\;  \mu_0 = (\al (1-\bty\cdot\ny))^{-1} . \ee{anis1}

\n
The second condition (cf. \rf{lafac}) gives the explicit form of the function \( \mu_0: \GC_0 \ra \R^+ \).

Now, given that $\La$ belongs to $\LCU$, the second equation follows from the first one and from the fact that $\La l$ must be future lightlike (since $l$ is); hence:

\[ |A(\ny - \bty)|^2 - ( \al (1-\bty\cdot\ny))^2 = 0. \]

\n
Thus \rf{anis1} can be reduced to three scalar equations

\be ( \al (1-\bty\cdot\ny) )^{-1} (A(\ny -\bty)) = \ny, \ee{lie}

\n
which, however, are themselves not independent, being under the constraint that the left hand side has  (3-dimensional Euclidean) norm equal to 1. It follows that the maximum number of independent conditions in \rf{anis1} on $\La$ is 2, thus $\GC_0$ (and therefore  also $\hGC_0$) is a 4-dimensional  Lie subgroup of $C\PCU^+$ (in fact from \rf{lie} it follows that $\GC_0$ is closed, and every closed subgroup of a Lie group is a Lie subgroup).\footnote{For the foundations of Lie group theory, see, for example, \cite{btd}; the cited theorem can be found in this reference at pp. 28-9.} Therefore the anisotropic Poincar\'e group is 8-dimensional, in contrast with the 10 dimensions of the Poincar\'e group(s), but in agreement with the classical case (\S 3.1). The fact that $\GC$ singles out a privileged direction in 3-space justifies considering it the structure group of a form of {\sl anisotropic special relativity} (ASR). As we shall see, it is not the only possible one.

 Let us investigate further the nature of $\GC_0$ and $\hGC_0$ as Lie groups. For every nonzero $\Vy$ in the open ball $B(\0, c)$ of $\R^3$ with radius $c$ we define ($\uy = \vers (\Vy)$)

\[ \my (\Vy) := \frac{\ny +((\al -1) (\uy\cdot\ny) -\al\bt)\uy}{\al (1-\bty\cdot\ny)}. \]

\n
This is the unit vector in $\R^3$ such that $\La (\Vy) l$ is proportional to $(\my (\Vy), 1/c)^T$, the proportionality factor being $1/\mu_0 (\La (\Vy))$ (compare with \rf{eilos}). Notice that $\my (\Vy)$ is never equal to the opposite of $\ny$, and that for every $\Vy$ proportional to $\ny$, and only in this case, one has $\my (\Vy) = \ny$. For $\Vy$ non-proportional to $\ny$, let $S(\Vy)$ be the rotation of $\R^3$ with axis $\my (\Vy)\we\ny$ mapping  $\my (\Vy)$ to $\ny$; we extend this definition by putting $S (\Vy) := I_3$  in the other cases.  Clearly the map $S: B(\0, c)\ra SO(3)$ which we have just defined is continuous. By construction the matrix \( \Si (\Vy)\La(\Vy) \), where $\Si (\Vy) = \Si_{S(\Vy)}$, is an element of $\GC_0$. It is easy to check that every element $X$ of $\GC_0$ can be expressed uniquely as the product 

\[ X = \Si_S\Si (\Vy) \La(\Vy) \]

\n
where $S$ is in the isotropy subgroup $H$ of $\ny$ in $SO(3)$. Thus the map:

\be F: B(\0, c)\times H \ra \GC_0, \; (\Vy, S) \mapsto \Si_S \Si (\Vy)\La(\Vy) \ee{omeo}

\n 
is bijective and, as a matricial product of three continuous maps, is itself continuous; it is clear that its inverse is also continuous, so $\GC_0$ is homeomorphic to $\R^3 \times SO(2)$, which is to be compared with the homeomorphism of $\LCU^+$ itself and $\R^3\times SO(3)$, based on the map

\[ B(\0, c) \times SO(3) \ra \LCU^+, \; (\Vy, S) \mapsto \Si_S \La (\Vy). \]  

The map $F$ makes it possible to find a basis for the tangent space at $I$ of $\GC_0$, which can be identified with the Lie algebra $L(\GC_0)$ of the group. A computation exploiting map \rf{omeo} shows that a basis of  $L(\GC_0)$ is given by the four matrices:

\be\si_0 =\left(\ba{cccc} 0 & 0 & 0 & 0 \\ 0 & 0 & -1 & 0\\0 & 1 & 0 & 0 \\ 0 & 0 & 0 & 0\ea\right),\;
\si_1 =\left(\ba{cccc} 0 & 0 & 0 & -c \\ 0 & 0 & 0 & 0\\0 & 0 & 0 & 0 \\ -\frac{1}{c} & 0 & 0 & 0\ea\right), \]
\[ \si_2 =\left(\ba{cccc} 0 & -1 & 0 & 0 \\ 1 & 0 & 0 & -c\\0 & 0 & 0 & 0 \\0 & -\frac{1}{c} & 0 & 0\ea\right),\;
\si_3 =\left(\ba{cccc} 0 & 0 & -1 & 0 \\ 0 & 0 & 0 & 0\\1 & 0 & 0 & -c \\ 0 & 0 & -\frac{1}{c} & 0\ea\right).\ee{liea}

\n
The Lie brackets of this basis can be obtained by direct computation as 

\be [\si_0,\si_1] = 0\; [\si_0 , \si_2] = \si_3, \;  [\si_0,\si_3] = -\si_2, \; [\si_2, \si_3] =0, \; [\si_1, \si_2] = -\si_2, \; [\si_1, \si_3] = -\si_3. \ee{bracket}

Now the group of orientation-preserving similitudes of the Euclidean plane can be identified with the 4-dimensional subgroup $SIM (2)$ of $GL(3,\R)$ of all matrices of the form

\be  \left(\ba{cc} rA & \by \\ [4pt] \0^T & 1\ea\right)\; \mbox{where}\; A\in SO(2), r\in R^+, \by\in R^2, 
\ee{sim2} 

\n
and it is easy to see that its Lie algebra has a natural set of 4 generators satisfying the commutator identities \rf{bracket}.

We can put together what we have found in a proposition (cf. \cite{b06}):

\begin{pro} The anisotropic Lorentz group $\GC_0$ for a given ray is a connected 4-dimensional Lie group isomorphic to $SIM (2)$, having its Lie algebra generated by a system of generators satisfying \rf{bracket}. \hfill$\Box$ \end{pro}  

This will prove important in the following section.

Proposition 3.2 implies that the anisotropic Poincar\'e group $\GC$ is isomorphic to $SIM(2) \rtimes \R^4$, a 8-dimensional Lie group (the same dimension of the anisotropic Galileo group) which is also denoted in the literature by $ISIM(2)$ (\cite{ggp}). 

\subsection{Generalization}

So far we have assumed that $\la \equiv 1$. However, the mere conservation of a certain space {\sl direction} does not require so much. The necessary and sufficient condition for the coordinate change to preserve ray $L$ is \rf{anis}, or equivalently

\be \mu\La l\propto_+ l , \ee{anis_b}

\n
and it is easy to verify that also this condition defines a subgroup of $C\LCU$, namely $\HC = \R^+ \GC_0$. This subgroup has the unwelcome property that the conformal factor $\mu$ is {\sl independent} of the Lorentz group factor $\La$ (this property is unwelcome since it allows for a freedom of choice in the units which at this level we normally want to avoid). However, the way we have succeeded in creating a link between $\mu$ and $\La$ -- that is, by substituting the special \rf{anisp} for the general \rf{anis} -- is not unique. By choosing {\sl any} group homomorphism $\la: \GC_0 \ra \R^+$, we obtain a subgroup of $\HC$ in the form

\[ \hGC_{0,\la} := \{ \mu_0 (\La) \la (\La)\La \; : \; \La\in\GC_0 \}. \]

A class of natural examples is provided by the remark that all nontrivial group homomorphisms from $\R^+$ to itself are of the form $h (x) = x^s$ where $s$ is any nonzero real number. So we can posit

\be \la (\La) := \mu_0 (\La)^{-s}. \ee{r_anis}

\n
This condition generates a family of subgroups of $\HC$ parametrized by $s\in \R$.  As a matter of fact it can be proved that {\sl all} $\hGC_{0,\la}$ are of this form, since the following assertion is true:

\begin{thm} All group homomorphisms from $\GC_0$ to $\R^+$ are of the form:

\be \la_s: \GC_0 \ra \R^+, \; \La \mapsto ( \al (1-\bty\cdot\ny) )^s . \ee{lam}\end{thm}

\n
{\bf Proof} We must prove that if $\la$ is an homomorphism from $\GC_0$ to $\R^+$, then there is a real number $s$ such that $\la = \la_s$. To do this, we will exploit the circumstance that $\GC_0$  and $\R^+$ are Lie groups and the proposition that two homomorphisms from a connected Lie group to any Lie group, with equal tangent maps at the unit element, are equal (see for instance \cite{btd}, p. 24). In order to apply this result we must prove that there is $s\in\R$ such that $\la_{\ast I} = (\la_s)_{\ast I}$, where $I$ is the identity 4-matrix. 

\[ \la_{\ast I}: L(\GC_0) \ra L(\R^+) \cong \R\]

\n
Notice that since \( \La (\bt\ny)l = \al(1-\bt)l \), it follows that $\La (\bt\ny)$ is in $\GC_0$ for all $\bt\in]-1,1[$.
We shall develop the computations for the case that $\ny = (1,0,0)$; clearly this does not detract from the generality of the result. We have

\[ \la_{\ast I}(\si_0) =0, \; \la_{\ast I} (\si_2) = -[\la_{\ast I}(\si_0), \la_{\ast I}(\si_3)] =0,\;
\la_{\ast I} (\si_3) = [\la_{\ast I}(\si_0), \la_{\ast I}(\si_2)] =0 .\]

\n
On the other hand $\la_{\ast I} (\si_1)$ can be computed as the velocity vector of $\bt\mapsto\la (\La(\bt\ey_1)$; now all homomorphisms $f$ from the one-dimensional subgroup \( \{\La(\bt\ey_1)\;:\; \bt\in]-1,1[\}\) to $\R^+$ are of the form

\[ f(\La(\bt\ey_1)) = (\al(1-\bty\cdot\ey_1))^s . \]

\n
From this it follows that

\[ \la_{\ast I}(\si_1) =  f_{\ast I}(\si_1) = \left.\frac{d}{d\bt}\right|_{\bt=0} (\al(1-\bty\cdot\ey_1))^s = s , \]

\n
as was to be shown. \hfill $\Box$

The main consequence of the theorem is that anisotropic special relativity (ASR) is in fact a {\sl family} of different theories, parametrized by an exponent. In fact from \(\mu \La l = \la l = \mu_0^{-s}l\) it follows that $\mu\mu_0^s\La l = l$, and therefore $\mu\mu_0^s = \mu_0$. Thus, after changing $s-1 \leadsto r$, we have the general transformation : 

\be x'= \mu_0 (\La)^{-r} \La x + b, \; \mbox{with}\; \La\in \GC_0 , b\in \R^4. \ee{aglt}

\begin{df} The {\bf r-anisotropic conformal Lorentz group} is

\[ \hGC_{0,r} = \{\mu_0 (\La)^{-r} \La \; : \; \La\in \GC_0\}, \]

\n
while the {\bf r-anisotropic Poincar\'e group} is $\hGC_r :=  \hGC_{0,r}\rtimes T(\R^4)$. A $\hGC_r$-space is also called a {\bf r-anisotropic Minkowski space-time}.\end{df}

Of course all groups $\hGC_{0,r}$ are isomorphic to $\GC_0 \cong SIM(2)$, and the groups $\hGC_r$ for different values of $r$ are all topologically equivalent with $\GC_{AG}$ (the anisotropic Galileo group); however they are not isomorphic, so they provide a nontrivial deformation of $ISIM(2)$, as explained in \cite{ggp}.\footnote{Its authors denote ${\hat{\cal G}}_r$ by $DISIMb(2)$, their $b$ being my $r$.}We shall see (\S 4) how parameter $r$ can be brought in principle to empirical determination.

Notice that the case $r=0$ does {\sl not} correspond to standard special relativity (in contrast with what is suggested in \cite{b}, p. 112, and \cite{b92}, p. 571, but not in a later paper by the same author, \cite{b06}), since the constraint on the privileged direction just cannot be made to disappear this way, as will be made even more clear in the next subsection.

\subsection{From Poincar\'e group to Very Special Relativity}

So far we have discussed spatial anisotropy by following the natural path of a reduction of the conformal Poincar\' e group. In this approach spatial anisotropy shows itself in i) a conformal factor $\mu$, ii) a dimensional decrease of the structure group (from 10 to 8). However it is also possible to begin with the Poincar\' e group (which can itself be conceived as a reduction of the conformal Poincar\' e group via spatial isotropy), and {\sl reduce it further} by requiring the invariance of a null vector, rather than merely of a null direction (as we have seen, the latter request would give us $\hGC_0$).  

In fact, as is easy to see, the condition $\La l = l$ defines a (normal) closed subgroup $\GC'_0$ of $\GC_0$, hence a Lie subgroup. From \rf{anis1} we see that the defining condition reduces to \( \mu_0 (\La) = 1\), or

\[ \al (1-\bty\cdot \ny) = 1, \] 

\n
that is, either $\bt =0$ or $\bty\neq \0$  and 

\[ \bt =\frac{2\cos\te}{1+\cos^2\te}, \]
 
\n
where $\te$ is the angle between $\Vy$ and $\bty$, so a relative nonzero velocity parallel or orthogonal to $\ny$ is not allowed between admissible css.  This means, in my view, that the resulting anisotropic group is not a reasonable one as the structure group of a general physical theory. Using the representation \rf{sim2} it is easy to see that $\GC'_0$ is isomorphic to the group of positive isometries of the Euclidean plane, $E(2)$.   

Summing up: 

\begin{pro} The subgroup $\GC'_0$ of  the proper orthochronous Lorentz group comprising all matrices having $l$ as eigenvector with eigenvalue 1 is a Lie subgroup of dimension 3, with Lie algebra generated by $\si_0, \si_2, \si_3$ (as defined in \rf{liea}), isomorphic to $E(2)$. No matrix in this subgroup can have a (nonzero) velocity orthogonal or parallel to $\ny$. \hfill $\Box$\end{pro}

However, this is not the only 3-dimensional subgroup of $\GC_0$ (up to isomorphism). Another one arises from the Lie subalgebra generated by $\si_1, \si_2, \si_3$, and is isomorphic to the subgroup of homotheties of the Euclidean plane corresponding under the representation \rf{sim2} to the group of matrices

\be  \left(\ba{cc} r I_2 & \by \\ [4pt] \0^T & 1\ea\right)\; \mbox{where}\; A\in SO(2), r\in R^+, \by\in R^2, 
\ee{hom2}

\n
which is the group $HOM(2)$ of the homotheties of the Euclidean plane. This subgroup makes more physical sense because no relative velocity (with module smaller than $c$, of course) is forbidden. The same is true for all the 3-dimensional subgroups corresponding to subalgebras of the form $<\si_1 + k \si_0, \si_2, \si_3>$ (but not for the subgroup of  $C\PCU^+$ locally isomorphic to $SL(2,\R)$).

The interest of theoretical physicists for these subgroups of the Lorentz group, and the corresponding subgroups of the Poincar\'e group obtained by taking their semi-direct product with $\R^4$, has been re-awakened\footnote{See Remark at the end of subsection 3.2.} by Cohen and Glashow's paper on Very Special Relativity (VSR), the theory giving space-time a $G$-structure, where $G = G_0 \rtimes \R^4$ and 

\[ G_0 = SIM(2), E(2), HOM(2), E(2)\cap HOM(2) = T(\R^2) \cong \R^2 . \] 

\n
The reason to introduce this variant of special relativity is that  it provides a fine theoretical tool to deal with violations of Lorentz symmetry, since VSR together with any of the operations P, T, CP (i.e. parity inversion, time reversal, charge conjugation combined with parity inversion) leads to the Lorentz group (\cite{cg}). Gibbons {et al.} (\cite{ggp}) have shown that the family $\hGC_r$ is the only physically acceptable nontrivial deformation of $\GC_0$, which can be seen as an alternative justification for the study of $\hGC_r$-space-times if one accepts VSR as a starting point. The model for this argument is the relationship between Minkowski space-time and deSitter/anti-deSitter spacetimes based on the circumstance that the (anti)deSitter isometry groups can be seen as deformations, uniquely, of the Poincar\'e group -- with the cosmological constant (or, equivalently, the radius of the universe) playing the role of $r$.\footnote{A rigorous proof can be found in \cite{ln}. It is worth mentioning that the first author to sketch this argument was L. Fantappi\`e in 1954 (\cite{fan}), whose ideas were developed during several decades by his disciple G. Arcidiacono (see for instance \cite{ar76}).}

\subsection{Anisotropic special Lorentz transformation (ASLT)}

To stress the properties of the anisotropic transformations, it is convenient to consider the transformation which in ASR corresponds to the special Lorentz transformation (SLT) in special relativity:

\be \left\{\ba{rcl} x'^1 &=& \dss  \frac{x^1 - vt}{\sqrt{1-\bt^2}}
\\[4pt] x'^2 &=& x^2 \\[4pt] x'^3 &=& x^3 \\[4pt]
t' &=& \dss \frac{t - vx^1/c^2}{\sqrt{1-\bt^2}},\ea\right. \ee{tsl}

\n
which can be re-written as

\[ x'= \La (\bt) x, \, \mbox{where} \,
\La(\bt) =  \left(\ba{cccc} 1/\sqrt{1-\bt^2} & 0 & 0 & -c\bt/\sqrt{1-\bt^2} \\ [4pt] 0 & 1 & 0 & 0 \\ [4pt]  0 & 0 & 1 & 0
\\ [4pt] -\bt/(c\sqrt{1-\bt^2}) & 0 & 0  & 1/\sqrt{1-\bt^2}\ea\right).    \]

It is important to emphasize that while in standard SR there exists a SLT for every direction of the velocity, this is by no means the case in ASR: we can find a one-parameter subgroup of {\sl special} transformations for one spatial direction only -- namely, the privileged one.

One can easily see that the only vector of the form $(\ny, 1/c)$ which is also an eigenvector of $\La (\bt)$ is the one with $\ny = (1,0,0)$. Since by \rf{anis1}

\[ \mu_0 (\La(\bt)) = \frac{1}{\al (1-\bty\cdot\ny)} = \frac{1}{\al(1-\bt)} = \sqrt{\frac{1+\bt}{1-\bt}}, \]

\n
we conclude that, if we choose $l = (\ny, 1/c)$ as giving the privileged spatial direction, the only {\sl anisotropic} special Lorentz transformation (ASLT) is, by specialization of \rf{aglt}:

\be x'= \mu_0 (\La (\bt))^{-r} \La(\bt) x, \ee{aslt}

\n
the inverse transformation being:

\[ x = \mu_0 (\La (-\bt))^{-r} \La(-\bt) x', \]

Equation \rf{aslt} can be written as the following system of four scalar equations (which was taken as a starting point by Bogoslovsky, cf. \cite{gb99}, p. 1384): 

\be \left\{\ba{rcl} x'^1 &=& \dss (\frac{1-\bt}{1+\bt})^{r/2} \, \frac{x^1 - vt}{\sqrt{1-\bt^2}} 
\\[4pt] x'^2 &=& \dss (\frac{1-\bt}{1+\bt})^{r/2}\, x^2 
\\[4pt] x'^3 &=& (\dss \frac{1-\bt}{1+\bt})^{r/2} \,  x^3 
\\[4pt] t' &=& (\dss \frac{1-\bt}{1+\bt})^{r/2} \,\frac{t - vx^1/c^2}{\sqrt{1-\bt^2}}.\ea\right. 
\ee{tsla}

\n
In a form which is easily proven to be equivalent (in light-cone coordinates), \rf{tsla} can be found in \cite{ggp} as equation (12). In the (physically unrealistic) case $r=1$ this set of equations reduces to:

\be \left\{\ba{rcl} x'^1 &=& \dss \frac{x^1 - vt}{1+\bt} 
\\[4pt] x'^2 &=& \dss \sqrt{\frac{1-\bt}{1+\bt}} \, x^2 
\\[4pt] x'^3 &=& \dss \sqrt{\frac{1-\bt}{1+\bt}} \,  x^3 
\\[4pt] t' &=& \dss \frac{t - vx^1/c^2}{1+\bt}.\ea\right. \ee{tsla1}

\n
Notice that \rf{tsla} reduces to the standard SLT when $r=0$, but, as pointed out in the previous section, this does not imply that the  structure group reduces to the (proper orthochronous) Poincar\'e group. However, recent estimates suggest for $r$ a very small upper bound, like $|r|< 10^{-26}$ (\cite{ljhrf86}, \cite{chlorw89}), so that a further analogy has been recognized (\cite{ggp}) between $r$ and the cosmological constant, since in both instances a major theoretical problem takes the form of the question: why should this parameter be so small? 

In the next section I will develop a kind of test of spatial isotropy which is completely independent of optics.

\section{Anisotropic proper time and the twin paradox}

The general expression for the r-anisotropic Lorentz transformation \rf{aglt} can be re-written in 3-dimensional terms as

\[ \left\{\ba{rcl} \ry' &=& \dss(\frac{1-\bty\cdot\ny}{(\sqrt{1-\bt^2})})^r A(\ry -t\Vy) + \by, \\ [5pt]
t' &=& \dss\frac{(1-\bty\cdot\ny)^r}{(\sqrt{1-\bt^2})^{r+1}}(t-\frac{\bty\cdot\ry}{c}) + b^4, \ea\right. \]

\n
where (cf. Appendix)

\[ A^TA = I_3 +\frac{\al^2}{c^2} \Vy\Vy^T, \; A(\ny-\bty) = \al (1-\bty\cdot\ny)\ny, \; \det A>0.\]

A process at rest in $\phi$ with proper duration $\De t$ will be observed by $\phi'$ as having the duration $\De t'$:

\[ \De t'=  \dss\frac{(1-\bty\cdot\ny)^r}{(\sqrt{1-\bt^2})^{r+1}}\De t, \]

\n
or equivalently

\[ \De t = \frac{\sqrt{1-\bt^2})^{r+1}}{(1-\bty\cdot\ny)^r} \De t' .\]

At the right-hand side we wish to express all in $\phi'$-quantities. Since $|\bty'| = |\bty|$ (see \S 5.1 on reciprocity), the only term which does not satisfy this condition is the denominator of the fraction.  However, $\mu_0$ is a group homomorphism, so we have

\[ \al (1-\bty\cdot\ny) = \mu_0 (\La)^{-1} = \mu_0 (\La^{-1}) = \frac{1}{\al (1-\bty'\cdot\ny)}, \]

\n
and therefore

\[ \De t = \frac{(1-\bty'\cdot \ny)^r}{(\sqrt{1-\bty'})^{r-1}}\De t' .\]

This formula allows us to state the {\sl Proper Time Principle} for ASR with privileged direction $l$ and exponent $r$: the proper time measured by a clock along a timelike worldline is given by

\be \hat{\tau} = \int^{t_2}_{t_1} \frac{(1-\bty\cdot \ny)^r}{(\sqrt{1-\bty})^{r-1}}dt, \ee{aptp} 

\n
where the quantities at the right-hand side are meant to be measured in any admissible coordinate system $\phi$ and $t_1$ and $t_2$ are the $\phi$-times of, respectively, the initial and the final event of the worldline. In invariant terms we can write (putting $\dot{x}: =dx/dt$)

\[ 1 -\bty\cdot\ny = \frac{1}{c}g (l, \dot{x}) \]

\n
where $g$ is the standard Lorentzian metric in $\R^4$, and therefore the integrand in \rf{aptp} can be written as the square root of 

\[ d\hat{\tau}^2 = \frac{1}{c^2} (\frac{g(l,\dot{x})^2}{|g(\dot{x},\dot{x})|})^r g(\dot{x}, \dot{x}) dt^2 \]

\n
which originates from the Finsler metric defined by Bogoslovsky

\be d\si^2 = c^2 d\hat{\tau}^2 = (\frac{g(l,dx)^2}{|g(dx,dx)|})^r g(dx, dx) . \ee{fins}

\n
The derivation provided here seems to me more satisfactory than those I found in the literature. It is easy to check directly that the metric \rf{fins} is $\hGC_r$-invariant; conversely, all 8-dimensional subgroups of $C\PCU^+$ which leave \rf{fins} invariant coincide with $\hGC_r$.

\begin{figure}[htp]
\centering
\includegraphics[totalheight=0.3\textheight ]{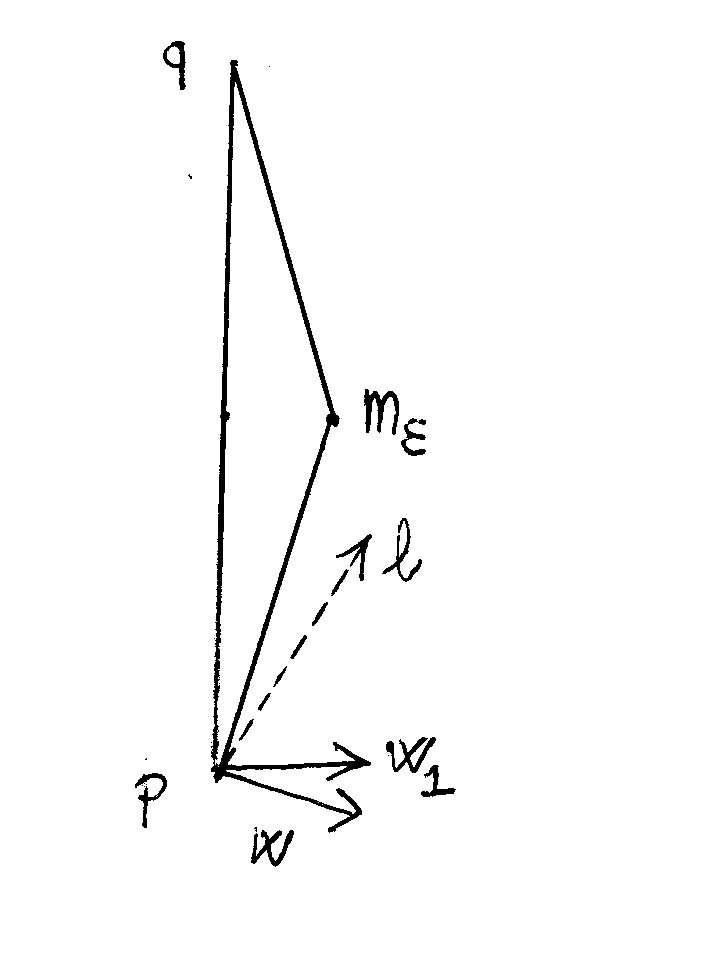}
\caption{Anisotropic `twin paradox'}
\end{figure}

Let us apply \rf{aptp} to compute the proper time from an event $p$ to an event $q$, chronologically following $p$. If $q-p$ is directly proportional to the inertial observer $u$, then there is a $b\in \R^+$ and a unit vector $w$ orthogonal to $u$ such that

\[ \ell = b (w + \frac{u}{c}) .\]

\n
Consider now any other unit vector $w_1 \perp u$, and put $\cos\te := g(w, w_1)$. Let the segment $pq$ be the worldline of the `stationary' twin, and let $m_\ep$ be the turning point of the `travelling' twin, where

\[ m_\ep = p +\frac{1}{2} (q-p) + c\ep w_1 ,\]

\n
with $0<\ep<\frac{a}{c}$ and $a:=|q-p|/2$, a condition ensuring that the broken line $p m_\ep \cup m_e q$ is a timelike worldline. A computation shows that the proper time measured by the first twin is:

\[ \hat{\tau}_I = 2\frac{ab^r }{c}, \]

\n
while for the other twin we obtain:

\[ \hat{\tau}_{II} = \dss\frac{(1-\frac{c\ep}{a}\cos\te)^r + (1+\frac{c\ep}{a}\cos\te)^r}{(1-\frac{c^2\ep^2}{a^2})^{r/2}}  \frac{ab^r}{c}\sqrt{1-\frac{c^2\ep^2}{a^2}} \]

\n
and therefore

\be\frac{\hat{\tau}_{II}}{\hat{\tau}_{I}} = \frac{1}{2} \dss((1-\frac{c\ep}{a}\cos\te)^r + (1+\frac{c\ep}{a}\cos\te)^r) (\sqrt{1-\frac{c^2\ep^2}{a^2}})^{1-r} \ee{diffag}

In \cite{b83} (p. 183) a simple general proof is given of the fact that $\hat{\tau}_I$ is a strict maximum among nearby worldlines, just as in Minkowski space-time. Formula \rf{diffag} shows how the effect of differential aging depends on the direction of the travel, and proves that the twin paradox provides in principle an experimental test of spatial isotropy, in agreement with \cite{b92} (p. 571). Of course in the exceptional case $r=0$ this test would not be adequate, and one should rely instead on subtler effects, for instance from particle physics (\S 3.6).

\section{Reciprocity, symmetry, light-speed isotropy}

In this section I shall deal with a number of conceptual issues related to the theory we have just developed, and that have often produced confusing statements in the literature. At the root lies, as we shall see, the wrong assumption that a ``principle of relativity" must contain all sorts of ``symmetries" between admissible coordinate systems. From an historical point of view it is interesting that so many authors for decades may have failed to focus certain important logical distinctions, probably because of the mesmerizing effect of the very word `relativity'. To anticipate two of our main claims: in special relativity, 1) the {\sl geometric} space relative to an inertial observer is Euclidean (even if one only accepts the two-way isotropy: see \cite{mm}, pp. 794-6), yet the {\sl physical} space may be anisotropic, and 2) the physical space may be anisotropic, yet the speed of light may be the same in all directions.

\subsection{Reciprocity principle}

Here is a standard formulation of the {\sl reciprocity principle}:

\begin{quote} \small
[\dots] the velocity of an inertial frame of reference $S$ with respect to another inertial frame of reference $S'$ is the opposite of the velocity of $S'$ with respect to $S$. [\cite{bg}, p. 1518; cf. \cite{lh}, pp. 172-3]
\end{quote}\normalsize

\n
This definition needs some explanations. First of all, in this form it makes sense only in the case of {\sl one} space dimension, since it corresponds to the very special equivalent conditions stated in proposition 2.3, 1), and which are not necessarily valid for an arbitrary Lorentz transformations.\footnote{In space-time terms, it is worth emphasizing that the ``velocity of $\phi$ with respect to $\phi'$'' and its reciprocal not only are not opposite to each other, but lie in different rest spaces!} On the other hand for all Lorentz matrices $\La$ it is true that $|A\Vy| = \al |\Vy|$ (an analogous statement holds for the Galileo group), so this is the correct form of the `reciprocity principle' in a general setting.  

In the article from which the preceding quotation has been extracted, the reciprocity principle is derived from ``the three basic postulates of the special theory of relativity" (and of classical physics as well!) ``namely, the principle of equivalence of inertial frames, the homogeneity of space-time, and the isotropy of space" (the invariance of the velocity of light is not assumed). Though legitimate in its own rights,\footnote{More on this will be said in a forthcoming paper.} results of this kind (of which scores have been published during the last hundred years) should not be taken to imply for the spatial isotropy condition a unique role in the validity of the reciprocity principle. In fact, since the reciprocity principle holds for $C\PCU^+$ (the conformal factor playing no role in the computation of the matrix velocities), it also holds for all its subgroups, including $\hGC_{0,r}$, for every $r$. 

\subsection{Symmetry of relativistic effects}

There is a naive criticism of relativity, which says that the length contraction and time dilation are `asymmetric' effects in standard special relativity, and that this contradicts the principle of relativity. For instance, the length contraction formula is

\be L'= L\sqrt{1-\bt^2}, \ee{slc}

\n
and if one solves it with respect to $L$, then the resulting formula looks like there is a length {\sl dilation}, which in turn seems to contradict the ``equivalence" between the coordinate systems $\phi$ and $\phi'$.\footnote{A variant of this argument is contained in Herbert Dingle's ``Preface'' to \cite{d72}.}

At a basic level this specious objection admits an obvious, blunt reply:  \rf{slc} is not the right equation to use if we wish to find out what happens when $\phi$ and $\phi'$ are swapped (in fact \rf{slc} assumes that the rod lies at rest in $\phi$, so its {\sl initial conditions} are not symmetric with respect to $\phi$ and $\phi'$). The same reply at the same level holds for the criticism of the seeming `asymmetry' of the time dilation formula. In a famous 1921 book Arthur S. Eddington described the length contraction effect by emphasizing its ``reciprocity". The relevant passage deserves to be quoted at length:

\begin{quote}\small
{\sl It is the reciprocity of these appearances -- that each party should think the other has contracted -- that is so difficult to realize}. Here is a paradox beyond even the imagination of Dean Swift. Gulliver regarded the Lilliputians as a race of dwarfs, and the Lilliputians regarded Gulliver as a giant. That is natural. If the Lilliputians had appeared dwarfs to Gulliver, and Gulliver had appeared a dwarf to the Lilliputians -- but no! that is too absurd for fiction, and is an idea only to be found in the sober pages of science.

{\sl This reciprocity is easily seen to be a necessary consequence of the Principle of Relativity}. The aviator must detect a FitzGerald contraction of objects moving rapidly relatively to him, just as we detect the contraction of objects moving relatively to us, and as an observer at rest in the aether detects the contraction of objects moving relatively to the aether. {\sl Any other result would indicate an observable effect due to his own motion through the aether}.

Which is right? Are we or the aviator? Or are both victims of illusions? It is not illusion in the ordinary sense, because the impressions of both would be confirmed by every physical test or scientific calculation suggested. No one knows which is right. No one will ever know, because we can never find out which, if either, is truly at rest in the aether.

It is not only in space but in time that these strange variations occur. [\cite{ed}, pp. 23-4; italics added]
\end{quote}\normalsize

The anisotropic theory we have discussed shows that this reply is not entirely correct, and Eddington's remarks on the ``reciprocity of appearances" are plain wrong. In anisotropic special relativity two arbitrarily chosen admissible coordinate systems are {\sl not} symmetric in the sense described by Eddington. Let us consider the length deformation effect in the two cases for $\phi$ and $\phi'$ linked by an ASLT as in \rf{tsla} with $\bt>0$ and $r >0$. If the rod is at rest in $\phi'$ along the $x'^1$-axis, then we have for the length measured by $\phi$ with respect to the proper length $L'_0$

\be L = (\frac{1+\bt}{1-\bt})^{1/2}\sqrt{1-\bt^2} L'_0 = ((1+\bt)^{1+r} (1-\bt)^{1-r})^{1/2} L'_0 =  :f_1 (\bt) L'_0, \ee{con1}

\n
while for a rod at rest in $\phi$ along the $x^1$-axis we have that the length $L'$ measured by $\phi'$ with respect to the proper length $L_0$ is

\be L' = ((1-\bt)^{1+r} (1+\bt)^{1-r})^{1/2} L_0 =:  f_2 (\bt) L_0 . \ee{con2}

Notice that since 

\be \frac{f_1 (\bt)}{f_2 (\bt)} = (\frac{1+\bt}{1-\bt})^r > 1, \; f_1 (\bt) f_2 (\bt) = 1-\bt^2 <1 \ee{conineq}  

\n
we have that \rf{con2} is always a `contraction'. On the other hand, 

\[ f_1'(\bt) = \frac{1}{f_1 (\bt)}(\frac{1+\bt}{1-\bt})^r (r-\bt) \] 

\n
thus

\[f'_1 (\bt)>0  \Longleftrightarrow  \bt < r\]

\n
and since $f_1 (0) = 1$ we have that \rf{con1} means that for an interval of values of $\bt$ there is a `dilatation' (and in the case $r=1$ it is {\sl always} a `dilatation'!). In other terms, there is not even {\sl qualitative} symmetry, and yet this fact does {\sl not} provide in itself ``an observable effect due to [$\ldots$] motion through the aether". It follows that the symmetry of effects is not a consequence of the principle of relativity {\sl per se}, as several authors assumed, both among the supporters and the critics of special relativity, but a consequence of a supplementary postulate putting a constraint on the conformal factor.\footnote{Probably this mistake was favoured by the link between `relativity' and `symmetry' which is at the forefront of Einstein's relativity paper (\cite{ei05}); for a detailed analysis of this issue I refer to \cite{bm}. The whole `Dingle affair' can be considered as an outgrowth of the unduly expanded meaning given to `relativity' in the interpretation of special or even {\sl general} relativity.} In fact from the first of the \rf{conineq} it follows that assuming the symmetry of the length deformation is {\sl equivalent}  to decreeing $r=0$.  

A similar argument proves that the assumption of the symmetry of time dilatation leads to the very same conclusion, since the `reciprocal' time tranformations:

\[ t'= \mu_0^{-r} \al (t-\frac{1}{c}\bty\cdot\ry), \; t= \mu_0^{r} \al (t'-\frac{1}{c}\bty'\cdot\ry'),\] 

\n
imply the following formulas for the transformation of time intervals: 

\[ \De t' = \mu_0^{-r} \al \De t, \; \De t = \mu_0^{r} \al \De t, \]

\n 
for identical processes at rest, respectively, in $\phi$ and $\phi'$; the factors are always equal if and only if $\mu_0^r  \equiv 1$, that is $r=0$ (since $\bt >0$). We conclude that {\sl symmetry of relativistic effects in the special transformations is equivalent to spatial isotropy} -- while, as we have seen (\S 5.1) , the reciprocity principle is not.

\subsection{Spatial isotropy versus light-speed isotropy}

For all transformations in the conformal Poincar\'e group the speed of light is always $c$ in all directions, so {\sl spatial isotropy is not to be confused with the isotropy of the one-way velocity of light}: to obtain the latter we can choose  indifferently \rf{tsl} as well as \rf{tsla}. Indeed the basic requirement for light-speed isotropy is satisfied by any of the {\sl r-anisotropic} Poincar\'e groups, rather than by the standard one only. Thus, for instance, the Michelson-Morley experiment gives the same outcome also according to ASR, a fact that is either ignored or insufficiently stressed by most authors.\footnote{Including myself, see \cite{mp94}, p. 897, where `conformal Lorentz invariance' would have been more exact than just `Lorentz invariance'; an exception is the short paper \cite{cg}.} For this reason, experimental claims that `space is anisotropic' in the sense of a positive reinterpretation of the residual effects in Michelson-Morley-Miller type experiments (as in \cite{al97}) would falsify ASR as well as standard SR.

As is well known, the light-speed (an)isotropy is related to the so-called ``conventionality of simultaneity" which, however, is much less `conventional' than is commonly believed: as I have argued at length, if one accepts the `non-conventional' principles of special relativity (including, crucially, the proper time principle), then there is no methodologically defensible decision for the Reichenbach function $\ep$ other than choosing it as constant (the only possible constant value being the ``standard" $\ep = 1/2$) (\cite{mm}).

A statement like ``The principle of spatial isotropy is invoked on setting the $\ep$ of Einstein's general radar rule equal to $1/2$ [$\ldots$]'' (\cite{lh}, p. 174) would be more satisfactory by substituting ``is'' with `may be', but it is true that spatial isotropy collapses the theory into classical physics or special relativity, with its $\ep \equiv 1/2$. In any case, as we have seen, there is no doubt that spatial isotropy is experimentally testable even if one insists that the isotropy of the one-way velocity of light is not.

\section{Appendix}

We shall deal here with  the matrix formalism of the Lorentz group used in the main text. Every matrix $\La$ of the proper orthochronous Lorentz group $\LCU^+$ can be expressed in the form (cf. \rf{blocks}):

\be \La = \left(\ba{cc} A & -c A\bty \\ -\frac{\al}{c}\bty^T & \al \ea\right), \ee{app1}
\n
where $\al = (1-\bt^2)^{-1/2}$, $\bt = |\bty|$, and if $x'= \La x$, then $\bty = \Vy/c$, with $\Vy$ the 3-velocity of the primed coordinate system with respect to the unprimed one; we shall also say that $\Vy$ is the `velocity of $\La$' and denote it by $\Vy_\La$. The $3\times 3$ matrix $A$ satisfies the identity:

\be A^T A = I_3 + \al^2\bty\bty^T ,\ee{app2}

\n
which shows that $A$ is determined up to left multiplication by a matrix in $SO(3)$. 

If $\Vy$ is any vector in $\R^3$ with module strictly smaller than $c$, the SLT with velocity $\Vy$ is given by:

\[ \La (\bty) = \left(\ba{cc} I_3 + (\al -1)\uy\uy^T  & -c \al\bty \\ -\frac{\al}{c}\bty^T & \al \ea\right), \]

\n
where $\uy$ is the unit vector in the direction of $\Vy$, if defined; otherwise, if $\Vy =0$, also $\uy: =0$. It follows that every matrix in $\LCU^+$ is of the form $\La = \Si_S\La (\Vy)$ with $\Vy = \Vy_\La$, and where $\Si_S\in SO_4 (3)$, the group of spatial rotations of $\R^4$ defined as follows:

\[ SO_4 (3) = \{\Si_S \equiv\left(\ba{cc} S & \0 \\ [4pt] \0^T & 1\ea\right)\; : S\in SO(3)\}. \]

\n
{\bf Acknowledgment} I wish to thank two anonymous reviewers for references and remarks which led to improvements in the original text.

\end{document}